\documentclass[pra,superscriptaddress,notitlepage]{revtex4-1}

\usepackage{graphicx}
\usepackage{amsmath}
\usepackage{hyperref}

\def\PPI{P_2}		
\def\PPII{{P'_2}}	
\def\PPJ{P_1}		

\begin{document}

\title{Josephson bifurcation readout: beyond the monochromatic approximation}

\author{Yuriy~Makhlin}
\affiliation{Condensed-matter physics Laboratory, HSE University, Moscow, Russia}
\affiliation{Landau Institute for Theoretical Physics, acad.~Semyonov av. 1a, Chernogolovka, Russia}
\author{Alexander~B.~Zorin}
\affiliation{Department of Physics, Moscow State University, Moscow, Russia}

\begin{abstract}
We analyze properties of bifurcation quantum detectors based on weakly nonlinear superconducting resonance circuits, in particular, with application to quantum readout.
The developed quantitative description demonstrates strong influence of higher harmonics on their characteristics.
While this effect is relevant for various circuits,
including the conventional Josephson bifurcation amplifier and the parametrically driven circuit, we first focus on the period-doubling bifurcation under a force driving.
This kind of bifurcation is due to nominally quadratic nonlinearity, which enables parametric down-conversion of the driving signal at nearly double resonance frequency
to the basic mode. We analyze the effect of higher harmonics on the dynamics of the basic mode, inherent in a nonlinear
circuit, which in our case is based on a Josephson junction with a sinusoidal current-phase relation as the origin of nonlinearity.
We demonstrate that effects beyond the monochromatic approximation significantly modify the bare characteristics and evaluate their contribution.
Due to high sensitivity of this circuit to small variations of parameters, it can serve as an efficient detector
of the quantum state of superconducting qubits.
\end{abstract}
\maketitle

\section{Introduction}

Further development of high-efficiency quantum detectors for solid-state quantum-information circuits, in particular, for Josephson quantum bits, is a task of high importance,
relevant for realization of novel devices and investigation of quantum behavior of such circuits. Due to their high sensitivity and weak backaction, detectors based on bifurcation phenomena are extensively used (see, e.g., Ref.~\cite{Vijay} for review). These circuits are generally based on a bistability between low- and high-amplitude forced oscillations at a frequency close to the basic frequency of a resonator with a cubic nonlinearity (equivalently, Kerr or $\chi^{(3)}$-nonlinearity~\cite{RWBoyd}).
The dynamics of this circuit is described by the Duffing equation~\cite{LLMech}.
A device using this method of operation was first proposed and developed by Siddiqi et al.~\cite{Siddiqi-2004}.
In this device, known as a Josephson bifurcation amplifier (JBA), driving near the resonance frequency is used to induce oscillations, and at sufficiently strong driving the circuit bifurcates from a single-valued to a bistable regime. These two possible oscillation states have different amplitudes and phases, but the same frequency, equal to the drive frequency. When biased near the bifurcation point, this circuit is extremely sensitive to small changes of its parameters, especially its resonance frequency, which in its turn depends on the state of a coupled qubit. This device and its modifications, operating as threshold detectors, were investigated and used by various experimental groups~\cite{Siddiqi-2006,Boulant-2007,Mallet-2009,Kakuyanagi2013,Schmitt2014,Boutin_2021,Metcalfe2007}.

A different threshold detector~\cite{pdbr}, termed a period-doubling bifurcation readout (PDBR), is based on a parametric period-doubling bifurcation in an externally driven Josephson-junction-based resonator with a quadratic nonlinearity (also known as non-centrosymmetric or $\chi^{(2)}$-nonlinearity~\cite{RWBoyd}).
A bistability in this circuit is developed between the zero state and an oscillation state at the basic frequency, whereas the drive is applied at a double frequency.
A quadratic nonlinearity in the current-phase relation $I(\varphi)$ of the Josephson junction is provided by a dc current bias, $|I_0|<I_c$, where $I_c$ is the critical current (see Fig.~\ref{fig:EqvSchm}a). Alternatively, such nonlinearity can be obtained using a constant-flux biased rf-SQUID~\cite{Zorin-JTWPA16}, as shown in Fig.~\ref{fig:EqvSchm}b, or an asymmetric multi-junction SQUID or the so called SNAIL circuit~\cite{Frattini2017}. This PDBR regime was experimentally realized in a microwave-driven superconducting Nb CPW resonator including an rf-SQUID~\cite{PorschThesis}.
A similar approach was used in a demonstration of a SNAIL cavity-based parametric amplifier with a large dynamic range~\cite{Sivak2019}.

Instead of nonlinearity-assisted pumping of the circuit, a period-doubling bifurcation may arise due to parametric modulation of the inductance or capacitance.
This regime was studied by Dykman et al.~\cite{Dykman-98,Dykman08} and demonstrated experimentally in a superconducting coplanar waveguide cavity including a magnetic-flux modulated dc-SQUID by Wilson et al.~\cite{Wilson}
Eventually, a parameter-modulated nonlinear circuit can also be used to detect quantum states of Josephson qubits; as we show below, its analysis is similar and its properties are comparable to those presented earlier in Ref.~\cite{pdbr}.
Since such readout strategy also implies generation of oscillations at the half-frequency of the drive, below we refer to this kind of device as PDBR-2, while PDBR-1 is reserved for the PDBR with quadratic nonlinearity and a force drive~\cite{pdbr}.
All these bifurcation-based devices could be used as efficient quantum detectors, and it is important to accurately analyze their behavior.

To describe the behavior of such a detector, one needs to find stationary states of a driven system and then analyze their stability and relaxation toward stable states.
This analysis requires a quantitative description of the device dynamics. For the simplest description of the circuit, one normally assumes that only oscillations at the basic frequency $\omega$ are induced (on top of a weak linear response at the drive frequency $2\omega$ in the case of PDBR-1), cf.~the literature cited above. To find the amplitude of these oscillations, one retains only the first harmonic of the evolution equation.
This approach was referred to as `monochromatic approximation' in Ref.~\cite{IthierThesis} in the case of JBA, and we use this term also for the case of the drive at the double frequency (although in this case it also involves the `trivial' weak response at $2\omega$ mentioned above).

Here we demonstrate that the monochromatic approximation is not sufficiently accurate, in the following sense:
in the effective equation of motion (EOM) for (the amplitude of) the basic harmonic the monochromatic approximation correctly yields the leading, linear term.
For the PDBR, this term determines the parameter range, where the zero solution becomes unstable, and where the quantum measurement is at all possible. For the JBA, the linear term defines the response in the linear regime.
However, subleading nonlinear terms govern development of the parametric instability in the PDBR regime as well as the full response for the JBA in the readout regime of interest.
The monochromatic approximation, as we demonstrate, fails to provide accurate values of the subleading terms.
We emphasize that the deviation is substantial, typically by a factor of order one, but in general it depends on the parameters of the effective potential of our nonlinear system and could be even stronger.
Obviously, this modification of the nonlinear term needs to be accounted for in the description of the device operation.
Such effects may generally appear in driven nonlinear resonators~\cite{LLMech}, and we develop a systematic approach to their quantitative description.

The effect considered here appears in generic nonlinear oscillators, but in our description below we have in mind circuits with a nonlinear Josephson inductance. The regimes of interest (i.e., JBA, PDBR-1, and PDBR-2) are described by similar equations, which differ only by the driving term. After setting the problem, we describe its solution for the PDBR-1 and summarize our results for the other two cases. We extend the general method of analysis of anharmonic oscillations~\cite{LLMech} to  account for various driving terms and higher nonlinearities.
We also note that recently effects of higher nonlinearities and multiple harmonics attracted attention in the field of Josephson-junction-based circuits. For instance, Ref.~\cite{KochetovFedorov} analyzed their effect on the gain saturation in a Josephson parametric amplifier in the regime below bistability, while the so called harmonic balance analysis for parametric amplifiers was presented in Ref.~\cite{Shiri2023}.

The circuits under consideration contain a phase-biased Josephson junction, which provides for parametric frequency conversion due to quadratic nonlinearity in the current-phase relation. When the circuit is current-driven at a frequency close to the double basic frequency of the resonator (the case of PDBR-1), a sharp onset of oscillations may occur at half of the drive frequency, i.e., close to the basic frequency.
The Kerr nonlinearity of the Josephson junction leads to a significant amplitude of these oscillations.
Exploiting this bistability between the zero state and finite-amplitude oscillations, one may efficiently detect small variations of the effective circuit capacitance or inductance, and therefore the quantum state of a coupled Josephson qubit.
The dynamics of the nonlinear Josephson circuit was analyzed in Ref.~\cite{pdbr} in a simplified model, where the renormalization corrections were introduced only at the final stage.
However, full characterization of this and similar devices and their potential use for quantum-state detection requires an accurate quantitative description of its behavior. Here we investigate an important correction and its influence on the dynamics of the circuit. Let us first briefly indicate the origin of this correction.

Driving at frequency $2\omega$, close to the double basic frequency of the resonator with an ordinary quadratic nonlinearity, induces oscillations at frequency $\omega$, close to its basic frequency. However, due to the inherent Kerr nonlinearity unaccounted oscillations at multiples $n\omega$ of this frequency are also induced, where $n = 0$, 3, 4, 5,\dots
In this article we analyze the feedback of these higher harmonics to the dynamics of the basic oscillations in the resonator. We derive the corresponding dynamic equations and show that their functional form is not modified, whereas the coefficients are changed, so we analytically find the contribution of the higher harmonics to these coefficients. We show that the non-dissipative part of the EOM is of Hamiltonian nature, which can be used to explain certain apparent coincidences of the coefficients and to find the most generic form of the low-order EOM describing small oscillations.

For the description of the circuit dynamics we use a version of the method of averaging or the method of slowly varying amplitudes, developed for the analysis of nonlinear systems (see, e.g., Refs.~\cite{Migulin,Stoker}). A specific extension of this approach was developed by Peter Kapitza for the analysis of his seminal Kapitza pendulum~\cite{Kapitza1951ZhETF,Kapitza1951UFN,KapitzaWorks} and is now part of the standard theoretical-physics toolbox (cf.~Ref.~\cite[\S~30]{LLMech}).
For example, Arnold describes~\cite{Arnold_Ponim} interesting connections between Kapitza's results and the Kolmogorov-Arnold-Moser theory (known as KAM theory) of quasi-periodic motions in Hamiltonian systems and their stability.

\section{Josephson bifurcation detectors and 
monochromatic approximation}

\begin{figure}
\begin{center}
\includegraphics[width=3.4in]{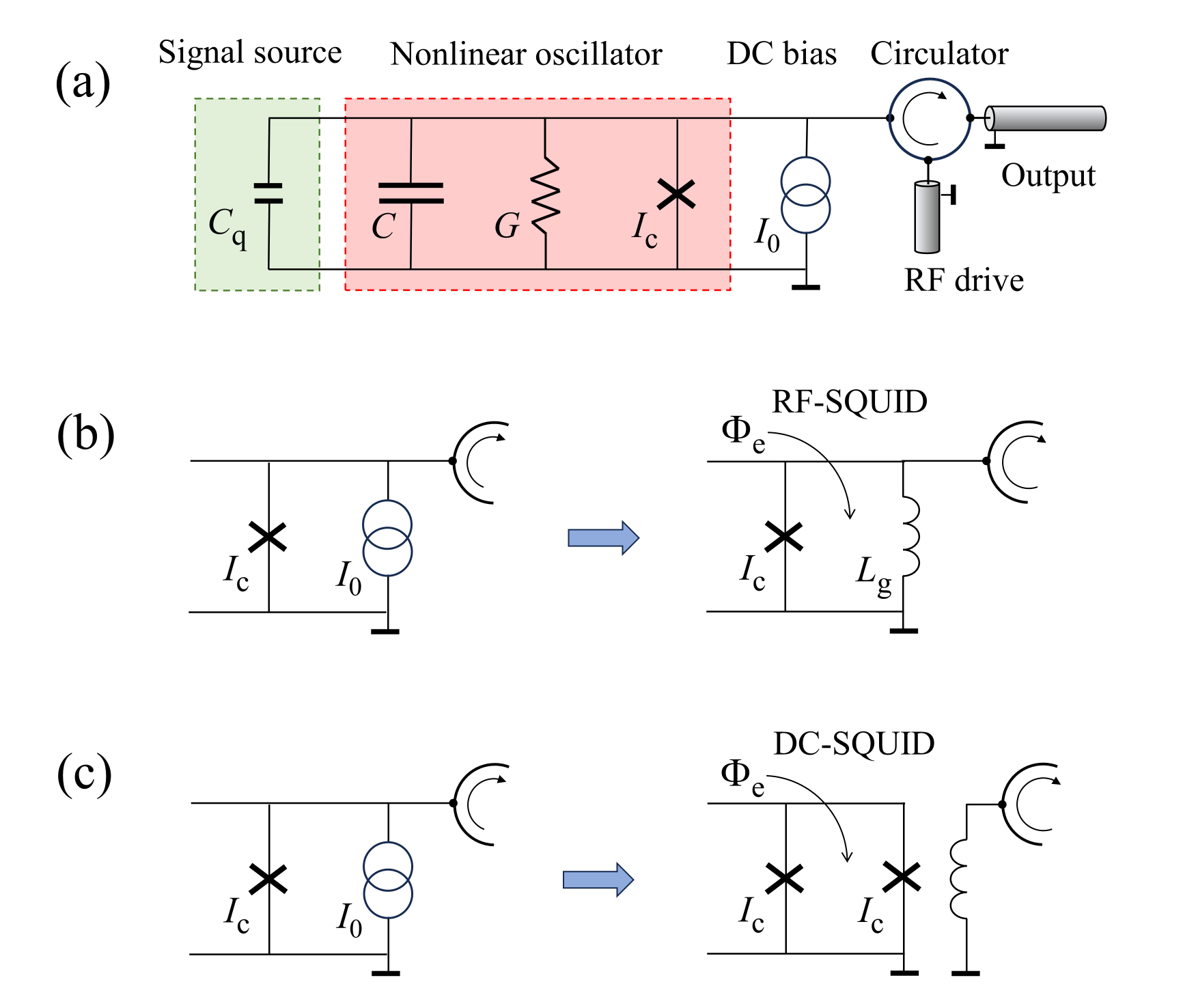}
\caption{(a) Electric diagram of the period-doubling bifurcation detector including a resonator formed by an inductance $L_J = \Phi_0/(2\pi I_c \cos\varphi_0) = L_{J0}/\cos\varphi_0$ of the Josephson junction with critical current $I_c$ (denoted by a cross symbol), externally biased by a dc current $I_0$,
and capacitance $C$, which includes the self-capacitance of this junction. Its effective conductance $G$ accounts for linear losses.
The resonator is coupled to a signal source, e.g., a charge qubit
with an effective quantum capacitance $C_q$, which depends on the state of the qubit~\cite{DutyQCapc2005,SillanpaaQCapac2005}. Both the microwave drive and readout are realized by means of a circulator.
(b) Modification of the generic circuit (a), where the current-biased Josephson junction is replaced by a flux-biased rf-SQUID
with a screening parameter $\beta_L = L_g/L_{J0} < 1$.
The drive at frequency $2\omega \approx 2\omega_p$ leads to the regime of period-doubling bifurcations (PDBR-1), while
the drive at frequency $\omega \approx \omega_p$ leads the conventional bifurcations (JBA) in both types of
circuits (a) and (b).
(c) The circuit based on a flux-driven symmetric dc-SQUID. For non-zero flux bias $\Phi_\textrm{e}$, the ac flux drive
at $2\omega \approx 2\omega_p$ may also lead to the period-doubling bifurcations due to periodic
modulation of the effective SQUID inductance (the PDBR-2 regime, see text).
\label{fig:EqvSchm}}
\end{center}
\end{figure}

The conventional circuit of a period-doubling-bifurcation detector proposed in Ref.~\cite{pdbr} PDBR-1 (see Fig.~\ref{fig:EqvSchm}a) consists of a dc-current-biased Josephson junction with critical current $I_c$, capacitance $C$ including both the self-capacitance of the junction and a possible external capacitance, linear conductance $G$, as well as an attached qubit, depicted here as a small capacitance $C_q$, presumably of quantum origin~\cite{SillanpaaQCapac2005,DutyQCapc2005}.
The circuit is driven by a harmonic signal $I_{\textrm{ac}}=I_A \cos 2\omega t$ at a frequency close to the double frequency of small-amplitude plasma oscillations $\omega_p$~\cite{KKLikharev-book}, i.e., $\omega \approx \omega_p$. The JBA can be represented by the same circuit with a different driving frequency, i.e. $I_{\textrm{ac}}=I_A \cos \omega t$.
A possible modification of this universal circuit is shown in Fig.~\ref{fig:EqvSchm}b, where the dc-current-biased junction is replaced by a constant-flux-biased rf-SQUID
with inductance $L_g$. In this case the effective linear inductance of the circuit $L$ is expressed via the phase-bias dependent Josephson-junction inductance $L_J$, i.e., $L^{-1}=L_J^{-1}+L^{-1}_g$. Nonlinear properties of such an element (quite similar to those of a dc-biased Josephson junction) are given by Eq.~(4) of Ref.~\cite{Zorin-JTWPA16}.
In the third case of PDBR-2 (see Fig.~\ref{fig:EqvSchm}c) the driving signal also has the double frequency but a different nature, viz., it modulates a reactance parameter of the resonator (the effective Josephson inductance of the dc-SQUID).

In the absence of fluctuations, the dynamics of the bare system (shown in Fig.~\ref{fig:EqvSchm}a, excluding the signal source, whose quantum state only slightly changes the plasma frequency $\omega_p$ of the entire circuit) is governed by the model of a resistively shunted Josephson junction~\cite{McCumber,Stewart}:
\begin{equation*}
\label{circuit-Eq}  C\frac{\Phi_0}{2\pi} \frac{d^{2}\varphi}{dt^{2}}
+ G \frac{\Phi_0}{2\pi} \frac{d\varphi}{dt}
+ I_c\:\sin\varphi = I_0+I_{\textrm{ac}}\,.
\end{equation*}
Here $\Phi_0 = h/2e$ is the magnetic flux quantum. A finite subcritical current bias, $|I_0|<I_c$, establishes
a dc phase drop $\varphi_0 = \arcsin(I_0/I_c)$ across the Josephson junction.
The small-signal expansion ($x = \varphi-\varphi_0 \ll 1$) of the Josephson supercurrent term includes the following dominant components: $\sin\varphi = \sin(\varphi_0+x) \approx \sin\varphi_0 (1-x^2/2+x^4/24)+\cos\varphi_0 (x-x^3/6)$. The resonant frequency of small oscillations of $\varphi$ around $\varphi_0$
is given by $\omega_p = (\cos\varphi_0)^{1/2}\omega_{p0}$~\cite{KKLikharev-book}, where $\omega_{p0} = (L_{J0}C)^{-1/2}$ is the bare plasma frequency and the Josephson inductance of the unbiased junction $L_{J0}=\Phi_0/(2\pi I_c)$.

After the small-signal expansion, the equation of motion for the dimensionless phase $x$ takes the form:
\begin{equation}\label{normalized_Eq}
\ddot{x} + x = \xi x -2\theta \dot{x} + \chi_2
x^2 + \chi_3 x^3 -\chi_4 x^4 + \textrm{driving term}\,,
\end{equation}
where the dots denote derivatives with respect to the
dimensionless time $\tau = \omega t$.
The dimensionless detuning and dissipation coefficients in Eq.~\eqref{normalized_Eq} are
\begin{equation}
\xi = 1-\bar\omega^{-2}\,, \ |\xi|\ll1\,,
\quad\textrm{and}\quad
\quad \theta=G/2\omega C \equiv 1/2Q \ll
1\,,\label{coeff:xitheta}
\end{equation}
respectively~\cite{pdbr}. The normalized frequency
\begin{equation}\label{kappa}
\bar\omega = \omega/\omega_p
\end{equation}
approaches one, $\bar\omega\to1$, on resonance.
Expressions for the nonlinear coupling coefficients,
$\chi_2=12\chi_4= \tan\varphi_0/2 \bar\omega^2$ and
$\chi_3 = 1/6\bar\omega^2$, are particular cases of the general expression~\eqref{coeff:betagammamu} below.

For the rf-SQUID configuration in Fig.~\ref{fig:EqvSchm}b, the resonant frequency and nonlinear coefficients can be expressed in terms of the dimensionless screening parameter 
$\beta_L = L_g/L_{J0}$~\cite{Tinkham-book,KKLikharev-book,Barone-Paterno-book,Clarke-Braginski-book} as
\begin{eqnarray}
&\omega_p^2 = \omega_{p0}^2 ( \cos \varphi_0 + \beta_L^{-1}) =  \ell^{-1}\omega_{p0}^2 \cos \varphi_0 \,,
\label{plasmafreq-rfSQUID}
&\\&
\chi_2 = 12\chi_4
= \ell \tan\varphi_0 / 2 \bar\omega^2  \,,
\quad \textrm{and} \quad
\quad  \chi_3 = \ell /6 \bar\omega^2 \,.
\label{coeff:betagammamu}
&\end{eqnarray}
Here a universal inductance renormalization factor $\ell$ is defined as
\begin{equation}
\ell =
\begin{cases}
1, &\textrm{(PDBR-1a)}\\
\displaystyle
\frac{\beta_L\cos\varphi_0}{1+\beta_L\cos\varphi_0}
= \frac{L_g}{L_g+L_J}= \frac{L}{L_J} \,,
&\textrm{(PDBR-1b)}
\label{def-lambda}
\end{cases}
\end{equation}
where $L_J(\varphi_0) = L_{J0}/\cos\varphi_0$.
With this definition, one can describe both PBDR-1a and PDBR-1b by the same Eqs.~\eqref{normalized_Eq}--\eqref{def-lambda}.
In particular, for PDBR-1b we find $\chi_2=12\chi_4= (1/2\bar\omega^2) \beta_L\sin\varphi_0 (1+\beta_L\cos\varphi_0)^{-1}$ and $\chi_3= (1/6\bar\omega^2) \beta_L\cos\varphi_0 (1+\beta_L\cos\varphi_0)^{-1}$.
As expected, the expressions for PDBR-1a can be obtained from those for PDBR-1b at $L_g\to\infty$, which yields $\beta_L\to\infty$ and $\ell\to1$.

In this rf-SQUID configuration, the dc phase bias $\varphi_0$ is set by an external magnetic flux $\Phi_e$. It can be found from the transcendental equation~\cite{KKLikharev-book} $\varphi_0 + \beta_L \sin \varphi_0 = 2\pi \Phi_e/\Phi_0$, which has a single solution for arbitrary $\Phi_e$ in the non-hysteretic regime $\beta_L<1$~\cite{Clarke-Braginski-book}, considered here. The external-flux dependence of the nonlinear coefficients $\chi_2$ and $\chi_3$ is shown, for instance, in Fig.~3 of Ref.~\cite{az-APL2021}.
Their values and signs can be efficiently controlled via $\Phi_e$ in the full range of phase bias, $-\pi \leq \varphi_0 < \pi$~(mod~$2\pi$), without any risk of the circuit switching from the superconducting to resistive state
(as in a stand-alone Josephson junction at $\varphi_0 \rightarrow \pi/2$~\cite{Tinkham-book}).
In particular, both the Kerr-free case, $\chi_2\neq 0$ with $\chi_3=0$ (reached at $\varphi_0 = \pm \pi/2$), and the pure
Kerr case, $\chi_3\neq 0$ with $\chi_2= 0$ (attained at $\varphi_0 = 0$ or $\pi$), are easily accessible
by tuning the flux $\Phi_e$.

Apart from the dynamics of PDBR-1, the generic nonlinear Eqs.~\eqref{normalized_Eq}--\eqref{def-lambda} can also describe the JBA. The respective driving terms read
\begin{align}
\textrm{driving term} &=&
3\PPI\,\cos2\tau, &\textrm{ where }\quad 3\PPI =
\frac{\ell}{\bar\omega^2 \cos \varphi_0} \frac{I_A}{I_c}\,,
&\textrm{ (PDBR-1a,b; current drive)}\label{eq:cdrivePDBR-1}\\
&&\PPJ\,\cos\tau, &\textrm{ where }\quad \phantom{3}\PPJ=
\frac{\ell}{\bar\omega^2 \cos \varphi_0} \frac{I_A}{I_c}\,,
&\textrm{ (JBA-a,b; current drive)}
\label{eq:cdriveJBA}
\end{align}
where the subscript of the drive amplitude $P$ indicates the drive frequency, and the factor 3 is introduced for convenience in further analysis.
The rf-SQUID resonator (see Fig.~\ref{fig:EqvSchm}b) can either be current-driven as above, or flux-driven by an additional ac flux applied to the SQUID loop, $\Phi_\textrm{ac}(t)=\Phi_A\cos2\tau$ (PDBR-1b)
or $\Phi_\textrm{ac}(t)=\Phi_A\cos\tau$ (JBA-b). In this case
\begin{align}
\textrm{driving term}
&=&3\PPI\,\cos2\tau, &\textrm{ where }\quad 3\PPI =
\frac{\ell}{\bar\omega^2 \cos \varphi_0}  \frac{2\pi}{\beta_L} \frac{\Phi_A}{\Phi_0}\,,
&\textrm{ (PDBR-1b; flux drive)}\label{eq:fdrivePDBR-1b}\\
&&\PPJ\,\cos\tau, &\textrm{ where }\quad \phantom{3}\PPJ=
\frac{\ell}{\bar\omega^2 \cos \varphi_0}  \frac{2\pi}{\beta_L} \frac{\Phi_A}{\Phi_0}\,.
&\textrm{ (JBA-b; flux drive)}
\label{eq:fdriveJBAb}
\end{align}

The evolution equation (\ref{normalized_Eq}), formally, can also describe the PDBR-2 circuit with a dc-SQUID shown in Fig.~\ref{fig:EqvSchm}c.
In contrast to PDBR-1 with forced oscillations, the frequency conversion in PDBR-2 occurs due to periodic modulation of the effective critical current, and hence inductance, of the dc-SQUID~\cite{Wilson,Migulin,Dykman-98}.
This modulation is controlled by an ac magnetic flux in the SQUID loop. Thus PDBR-2 can operate without quadratic nonlinearity, for instance, at zero current bias, $\varphi_0=0$.
An optimal dc flux bias $\Phi_e = \Phi_e^{\textrm{opt}} \approx \pm \Phi_0/3$ sets an effective critical current $I_c^\textrm{eff} = 2I_c |\cos (\pi\Phi_e/\Phi_0)| \approx I_c$ and allows one to combine efficient, nearly linear modulation of the critical current ($\partial I_c^\textrm{eff} / \partial\Phi_e\ne0$) with maximal possible swing.
Expanding the current through the dc-SQUID $I = I_c^\textrm{eff} \sin \varphi = I_c^\textrm{eff} (\varphi - \varphi^3/6 +...)$ at $\varphi_0=0$, we find the coefficients $\chi_3 = 1/6\bar\omega^2$ and $\chi_2=0$. An applied alternating flux $\Phi_{\textrm{ac}}(\tau) = \Phi_A \cos 2\tau$ modulates the critical current $I_c^\textrm{eff} = I_c^\textrm{eff}(\Phi_e^0)
[1+ f \Phi_\textrm{ac}(t)/\Phi_0]$ (and thereby the SQUID inductance).
At an optimal dc-flux bias point, $\pm\Phi_0/3$,
we find $f=(\partial I_c^{\textrm{eff}}/\partial \Phi_e)(\Phi_0/I_c^\textrm{eff}) \approx \mp\sqrt{3}\pi$.
The resulting driving term for PDBR-2 takes the form
\begin{equation}
\textrm{driving~term} = \PPII x\,\cos2\tau, \qquad \textrm{where}\qquad \PPII = \bar\omega^{-2} f \Phi_A/\Phi_0
\,.
\qquad\textrm{(PDBR-2)}
\label{eq:drivePDBR-2}
\end{equation}

In some realistic circuits (for instance, a flux-driven asymmetric dc-SQUID with different junctions, $I_{c1}\ne I_{c2}$), the applied ac flux can simultaneously induce both the force- and parametric-driving terms, i.e., $\PPI$ and $\PPII$ terms in Eqs.~\eqref{eq:fdrivePDBR-1b} and \eqref{eq:drivePDBR-2}, respectively.
Since the effective evolution equations \eqref{eq:A1tilde} and \eqref{normalized_Eq:par}, derived below respectively for PDBR-1 and PDBR-2, are of the same form, to the lowest order in the drive one can simply add the respective contributions with $\PPI$ and $\PPII$.

Thus Eq.~(\ref{normalized_Eq}) with appropriate driving terms allows us to consider both PDBR-1 and PDBR-2 circuits driven near the double resonant frequency and JBA circuits driven near the fundamental frequency.
Below we focus mostly on the PDBR-1 circuits, whereas our results can be easily extended to the case of JBA and PDBR-2.

\begin{figure}
\begin{center}
\includegraphics[width=3.3in]{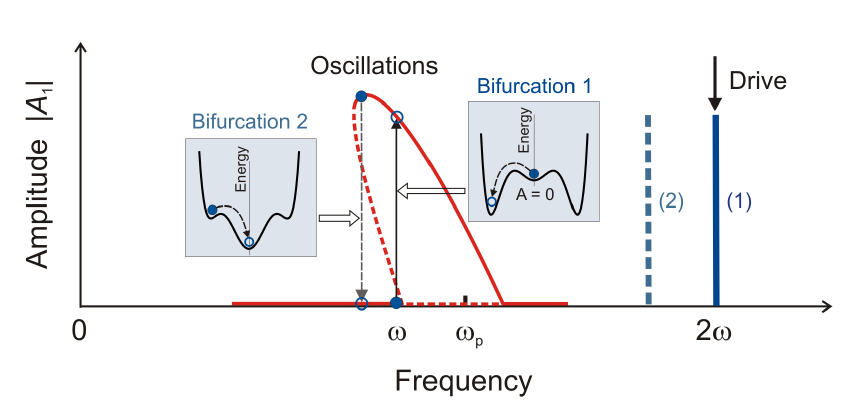}
\caption{Schematic plot of the parametric resonance curve (red): Amplitude $|A_1|$ of the fundamental-frequency Josephson-phase oscillations in PDBR-1 vs. frequency for a $2\omega$-drive signal with the drive amplitude above the bifurcation threshold.
The slope of the response curve is shown for positive Kerr nonlinearity, $\chi_3 > 0$. At $\chi_3 < 0$, the regime accessible in PDBR-1 with an rf-SQUID biased by flux
$\Phi_e$ not far from $\Phi_0/2$, the parametric-resonance curve has an opposite slope.
The dashed red line indicates an unstable state.
Increasing the the drive frequency induces a PDB (solid vertical arrow); when this frequency decreases, the system switches back to the zero state (dashed vertical arrow).
The insets schematically show transitions in the corresponding metapotentials~\cite{pdbr}.
Similar transitions may be induced by variation of other parameters, for instance, of the drive amplitude (which determines the width of the resonance peak), cf.~Ref.~\cite{pdbr}.
\label{fig:res-curve}}
\end{center}
\end{figure}

In the monochromatic approximation, one seeks a solution of Eq.~(\ref{normalized_Eq}) in the form $x \equiv
(A_1 e^{-i\tau} + \mathrm{c.c.}) - \PPI \cos 2\tau$.
Substituting this expression in Eq.~(\ref{normalized_Eq}),
one finds an evolution equation for the slow complex amplitude $A_1$ of the first harmonic, cf.~Ref.~\cite{pdbr},
where it was presented as a pair of equations for its absolute value $A$ and phase $\alpha$:
$A_1=\frac{1}{2}A e^{i\alpha}$. Rewriting them as a single equation for $A_1$, we thus find in the monochromatic approximation:
\begin{equation}
\label{eq:A1}
\dot A_1 =
-\left(\theta-i\frac{\xi}{2}\right) A_1
-\frac{i}{2}\chi_2 \PPI A_1^*
+\frac{3i}{2}\chi_3|A_1|^2 A_1
+i \chi_4 \PPI (3 |A_1|^2 A_1^* + A_1^3 )\,.
\end{equation}
However, this derivation of Eq.~\eqref{eq:A1} neglects possible deviations of the second harmonic from the pure drive $-\PPI\cos2\tau$ as well as contributions of the other harmonics, with numbers $n=0$ (dc) and $|n|>2$. The dynamics of these harmonics
(they all are loosely termed {\it higher harmonics} in this article) are coupled to the evolution of $A_1$.
We show below that their dynamics is faster than that governed by Eq.~\eqref{eq:A1},
and using separation of the time scales one can still derive
a closed equation of motion for $A_1$.
Since the effect of the higher harmonics arises due to the weak nonlinearity, it does not modify the leading, linear terms in the equation for amplitude $A_1$, but in general does modify the subleading terms.
The linear terms determine the stability range of the zero solution, $A_1=0$, under parametric pumping, $\PPI \neq 0$. However, the subleading, nonlinear terms are crucial for understanding the development of the instability.
Below we analyze the effect of the higher harmonics on the EOM (\ref{eq:A1}). Surprisingly, we find that
their effect does not modify the functional
form of this equation for $A_1$ but only modifies its coefficients.
This holds true also when one accounts for higher-order contributions in amplitude $\PPI$.

\begin{figure}
\begin{center}
\includegraphics[width=3.3in]{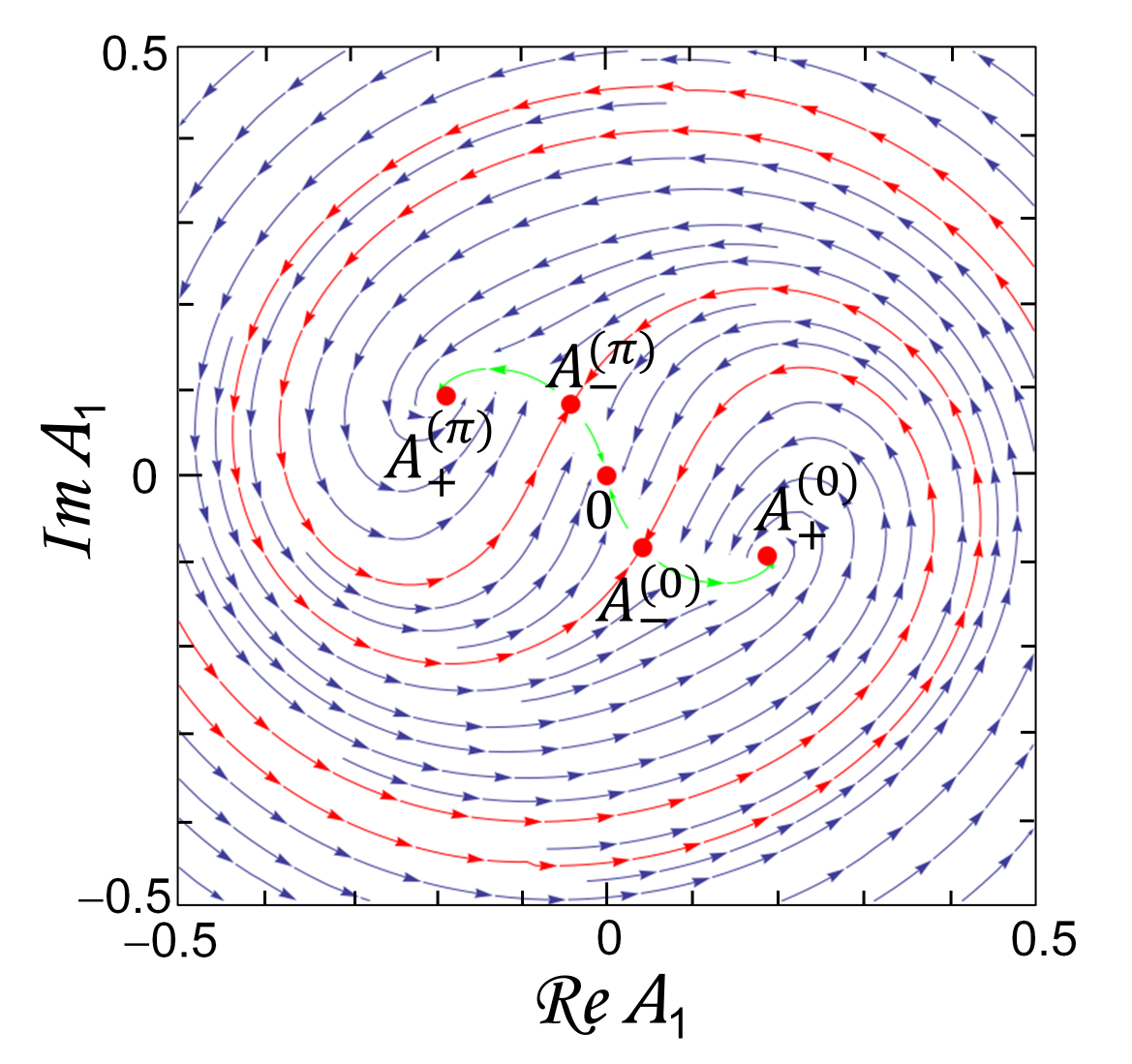}
\caption{Velocity field for the dynamics in the plane of the complex amplitude $A_1$, governed by Eq.~(\ref{eq:A1}) and shown for a circuit with the quality factor $Q=\omega C/G=25$, phase bias $\varphi_0=50^\circ$, and detuning $\xi=-0.045$.
The five red dots indicate stationary solutions and are arranged symmetrically:
the stable zero state, $A_1 = 0$, in the middle,
two degenerate non-zero states $A_+$ with a relative $\pi$-shift on the outside, and two degenerate $\pi$-shifted unstable solutions $A_-$ between them.
The red separatrix lines passing through the unstable solutions $A_-^{(\pi)}$ and $A_-^{(0)}$ delimit the basins of attraction for three stable states, $A_+^{(\pi)}$, 0, and $A_+^{(0)}$.
During the evolution, fast relaxation towards the bottom of the valley connecting all five stationary states (green line) is followed by subsequent slow dynamics along this green line~\cite{pdbr}.}
\label{fig:flow}
\end{center}
\end{figure}

An illustration of the flow in the phase space (the plane of the complex amplitude $A_1$) is shown in Fig.~\ref{fig:flow}.
Before we proceed with the analysis, we note that Eq.~(\ref{eq:A1})
allows one to find stationary solutions and analyze their stability (see Fig.~\ref{fig:res-curve}), as well as to study tunnel rates between the states in the bistable regime (illustrated in the insets). The latter are crucial quantities for the description
of the quantum measurement using this detector~\cite{pdbr}.

\section{Evolution beyond the monochromatic approximation}

The drive at frequency $2\omega$ generates, in the first place, oscillations at $2\omega$ (in the linear response) and $\omega$
(under parametric instability). Due to nonlinearities these oscillations induce also a weaker response at other multiple frequencies $n\omega$, $n=0, 3, 4, \dots$
This response should be self-consistently taken into account in order to derive the dynamics of the first harmonics in higher orders (due to nonlinear effects). In this section we perform this derivation.

Let us derive the EOM for the amplitudes of all harmonics,
induced by the $2\omega$-drive. We assume that the resulting oscillations are almost $2\pi$-periodic, and the phase variable can be represented in the following form:
\begin{equation}
x = \sum_{n=-\infty}^{+\infty} A_n(\tau) e^{-in\tau} \,,
\label{eq:modes}
\end{equation}
where the amplitudes $A_n(\tau)$ are slowly varying functions on the scale of the oscillation period, $\tau=2\pi$, so that the Fourier transform of $x(\tau)$ contains only frequencies close to integers.
Since $x$ is a real variable, the complex amplitudes $A_n$ satisfy the relation $A_{-n}=A_n^*$.

Below we derive the EOM for the complex amplitudes $A_n(\tau)$. We find that the dynamics of $A_{\pm1}$ is slower than that of the amplitudes with $n\neq \pm 1$, which allows us to find them as functions of $A_{\pm1}$ in the adiabatic approximation.
Thus we derive a closed EOM for complex amplitudes $A_{\pm1}$.

\subsection{Quasi-stationary values of higher harmonics}

To proceed with the derivation, we start with Eq.~\eqref{normalized_Eq} for $x$ in the form \eqref{eq:modes}.
Assuming small values of the parameters $\xi$, $\theta$, and $\PPI$ and a small amplitude of oscillations, $A\ll1$, we find that
\begin{equation}
(1-n^2) A_n - 2in\dot A_n + \ddot A_n
= (\xi +2in\theta)A_n
+ (\chi_2 x^2 + \chi_3 x^3 -\chi_4 x^4)_n
+ \frac{3}{2}\PPI\delta_{n,\pm2} \,,
\label{eq:n}
\end{equation}
where the last term on the rhs enters only the equation for
$n=\pm2$, and the subscript $n$ in the previous, higher-order term indicates that the $n$th term in the full Fourier series is taken (for instance, $(\chi_2 x^2)_n \equiv \chi_2 \sum_{k=-\infty}^\infty A_k A_{n-k}$).
This equation can be rewritten as a system of separate equations for each harmonic,
\begin{equation}
-2in\dot A_n + \ddot A_n = D_n(\{A_m\}), \quad n = 0, \pm 1, \pm 2...,
\label{eq:AnDn}
\end{equation}
where the rhs of \eqref{eq:AnDn} is given by
\begin{equation}
D_n(\{A_m\}) = (n^2-1+\xi +2in\theta)A_n + (\chi_2 x^2 + \chi_3 x^3 -\chi_4 x^4)_n + \frac{3}{2}\PPI\delta_{n,\pm2}\,,
\label{eq:Dn}
\end{equation}
and each $D_n$ depends on all harmonics $\{A_m\}$ with positive and negative $m$ due to nonlinear terms on the rhs.

Under the assumption that the amplitudes $A_n(\tau)$ are slow, the first term dominates on the lhs of Eq.~(\ref{eq:n}) for all harmonics with $n\ne\pm1$.
All these harmonics relax or average out sufficiently fast (as compared to the dynamics of $A_1$) to their instantaneous quasi-stationary values determined by $A_1$.
These values can be deduced from the conditions $D_n(\{A_m\})=0$ for all $n\ne\pm1$.
We remark that strictly speaking the higher harmonics evolve at frequencies of order $1$, which complicates separation into the modes in Eq.~(\ref{eq:modes}). However, when these higher
harmonics are fixed to their quasi-stationary values, their dynamics just slowly follow that of $A_1$, and that is sufficient for our purposes.

Thus, the amplitudes $A_{n\ne\pm1}$ as functions of $A_{\pm1}$ and $\PPI$ can be deduced from the condition that $D_n=0$ for all $n\ne\pm1$, which amounts to solving the  self-consistent equations
\begin{equation}
A_n = \frac{-1}{n^2-1+\xi +2in\theta} \left[ (\chi_2 x^2 + \chi_3 x^3 -\chi_4 x^4)_n + \frac{3}{2}\PPI\delta_{n,\pm2}
\right] \,.
\label{AnofA1}
\end{equation}
The resulting expressions for $A_{n\ne\pm1}$ in terms of $A_{\pm1}$ should then be substituted to Eq.~(\ref{eq:n}) with $n=1$ to find the EOM for the amplitude $A_1$ (and its conjugate $A_{-1}=A_1^*$).

\subsection{Hamiltonian form of EOM and gauge invariance}

Before proceeding with the derivation, let us remark that the final equations turn out to be Hamiltonian (except for the dissipative $\theta$-term,
like in Eq.~\eqref{eq:A1}). On one hand, the Hamiltonian form of the equation can be justified before the derivation; on the other hand,
it constrains the form of the final equation.

Indeed, one  can start from the Hamiltonian equations for position $x(t)$ and the corresponding momentum $p(t)$ and derive equations of motion for $A_1(\tau)$
by first changing $x,p$ for the harmonics $A_n$ and then integrating out all harmonics except $A_{\pm1}$.
This would give an effective Hamiltonian $H(A_1,A_1^*)$ for $A_1$ and the corresponding evolution equation,
\begin{equation}
\dot A_1 = \frac{i}{2} \frac{\partial H}{\partial A_1^*}\,.
\end{equation}

The fact, that $H$ is real, implies the following most general form of the lowest-order terms in the Hamiltonian:
\begin{eqnarray}
H(A_1,A_1^*) &=& \xi |A_1|^2 - \frac{\chi_2 \PPI}{2} \left[A_1^2+(A_1^*)^2\right]\nonumber\\
&+& \frac{3}{2}\chi_3 |A_1|^4
+ 2\chi_4 \PPI |A_1|^2 \left[A_1^2+(A_1^*)^2\right]
+ \nu \PPI^2 \left[A_1^4 + (A_1^*)^4\right]
\label{HamA1}
\,,
\end{eqnarray}
where $|A_1|^2 = A_1^*A_1$. The symmetry properties that we discuss in this section leave the coefficients in this Hamiltonian undetermined, and they need to be found by other means. However, our notations for these coefficients are in agreement with Eq.~\eqref{eq:A1}, see below.

Indeed, Eq.~\eqref{HamA1} includes all possible terms up to the 4-th order in $A_1$, which are real and `gauge invariant', i.e., invariant under a time translation. It implies that $A_1\to A_1 e^{i\psi}$, $A_1^*\to A_1^*e^{-i\psi}$, $\PPI\to \PPI e^{2i\psi}$
(since $\PPI$ is the second harmonic). Note, however, that all our equations are written in the gauge, where $\PPI$ is real,
whereas in general we could write, for instance, the second term on the rhs as $(\chi_2/2)\PPI^* A_1^2+{\rm h.c.}$

All the coefficients in Eq.~(\ref{HamA1}), $\xi$, $\chi_2$, $\chi_3$, $\chi_4$, and $\nu$ are functions of $\PPI$.
Because of the gauge invariance, the expansions in small $\PPI$ for the second and fourth terms (proportional to $\chi_2$ and $\chi_4$, respectively) begin with $\PPI$, while for the last term ($\propto\nu$) it begins with $\PPI^2$ --- hence it is small and appears only in higher orders. In fact, it does not appear in our low-order derivation, and  is irrelevant for weak pumping $\PPI$.

One can see that Hamiltonian (\ref{HamA1}), without the last, high-order term, gives rise to EOM of the form (\ref{eq:A1}).
So, the Hamiltonian nature of the equations and their `gauge invariance' imply that the corrections due to higher harmonics or higher-order terms in the pumping amplitude $\PPI$ do not change
the functional form of the equation for $A_1$. Moreover, they also explain why the last two terms in Eq.~(\ref{eq:A1})
include the same coefficients $\chi_4$: indeed, they stem from two conjugate components of the $\chi_4$-term in Hamiltonian (\ref{HamA1}).

It is noteworthy that the effective Hamiltonian\eqref{HamA1} for the basic harmonic $A_1$ is analogous to the effective potential for Kapitza's inverted pendulum~\cite{LLMech}.

\subsection{Corrections due to higher harmonics}

To find the coefficients in Hamiltonian \eqref{HamA1} to the leading order in $\PPI$, we solve the set of equations (\ref{AnofA1}).
To the lowest order in $A_1$ and $\PPI$ a few first harmonics are found to be:
\begin{eqnarray}
A_0 &=& 2\chi_2 |A_1|^2 + \left(\frac{1}{3}\chi_2^2-\frac{3}{2}\chi_3\right)
	[\PPI A_1^2 + \mathrm{c.c.}] +\dots \,,\label{eq:A0}\\
A_2 &=& -\frac{1}{2}\PPI -\frac{1}{3}\chi_2 A_1^2
	+ \left(\frac{7}{12}\chi_2^2+\chi_3\right) \PPI |A_1|^2 +\dots \,,\label{eq:A2}\\
A_3 &=& \frac{1}{8}\chi_2 \PPI A_1
	+ \left(\frac{1}{12}\chi_2^2 -\frac{1}{8}\chi_3\right)A_1^3 +\dots \,,\label{eq:A3}
\end{eqnarray}
where $\dots$ stands for terms of higher order in $A_1$.
In particular, the first term on the rhs of Eq.~\eqref{eq:A0} describes an additional drive-dependent phase offset due to rectification of harmonic oscillations by quadratic nonlinearity.
The harmonics $A_n$ of higher order $|n|>3$
do not contribute to the relevant terms of the EOM for $A_1$, and we further neglect corrections of order $\xi$ and $\theta$.
Substituting Eqs.~\eqref{eq:A0}--\eqref{eq:A3} into Eq.~(\ref{eq:n}) for $n=1$ and keeping only terms up to third order in $A_1$, we find the EOM for $A_1$ of the same form as Eq.~\eqref{eq:A1},
\begin{equation}
\label{eq:A1tilde}
\dot A_1 =
-\left(\theta-i\frac{\widetilde\xi}{2}\right) A_1
-\frac{i}{2}\widetilde\chi_2 \PPI A_1^*
+\frac{3i}{2}\widetilde\chi_3|A_1|^2 A_1
+i \widetilde\chi_4 \PPI (3 |A_1|^2 A_1^* + A_1^3 )\,,
\end{equation}
with the modified coefficients (marked by the tilde sign),
\begin{eqnarray}
\widetilde\chi_3 &=& \chi_3 + \frac{10}{9}\chi_2^2\,,\label{gtilde_fin}	\label{eq:gammatil}\\
\widetilde\chi_4 &=& \chi_4 -\frac{15}{16}\chi_2\chi_3 + \frac{7}{24}\chi_2^3 \,,	\label{eq:mutil}
\end{eqnarray}
while $\widetilde\xi=\xi$ and $\widetilde\chi_2=\chi_2$ to the lowest order in $\PPI$.
Using expressions (\ref{coeff:betagammamu}) for $\chi_2$, $\chi_3$, and $\chi_4$ via the phase bias $\varphi_0$, we find:
\begin{eqnarray}
\widetilde\chi_3 &\approx& \frac{\ell}{6} + \frac{5}{18} (\ell\tan\varphi_0)^2,\\
\widetilde\chi_4 &\approx& \frac{\ell\tan\varphi_0}{192}
[7 (\ell\tan\varphi_0)^2-15\ell+8] \,,
\end{eqnarray}
since $\bar\omega \approx 1$ near resonance.
Note that Eq.~(\ref{gtilde_fin}) is in agreement with the expression derived by Nayfeh~\cite{Nayfeh}.
According to Landau and Lifshitz~\cite{LLMech} this result can be interpreted as an extra correction to the resonant frequency of the anharmonic oscillator due to quadratic nonlinearity.

\begin{figure}
\begin{center}
\includegraphics[width=3.4in]{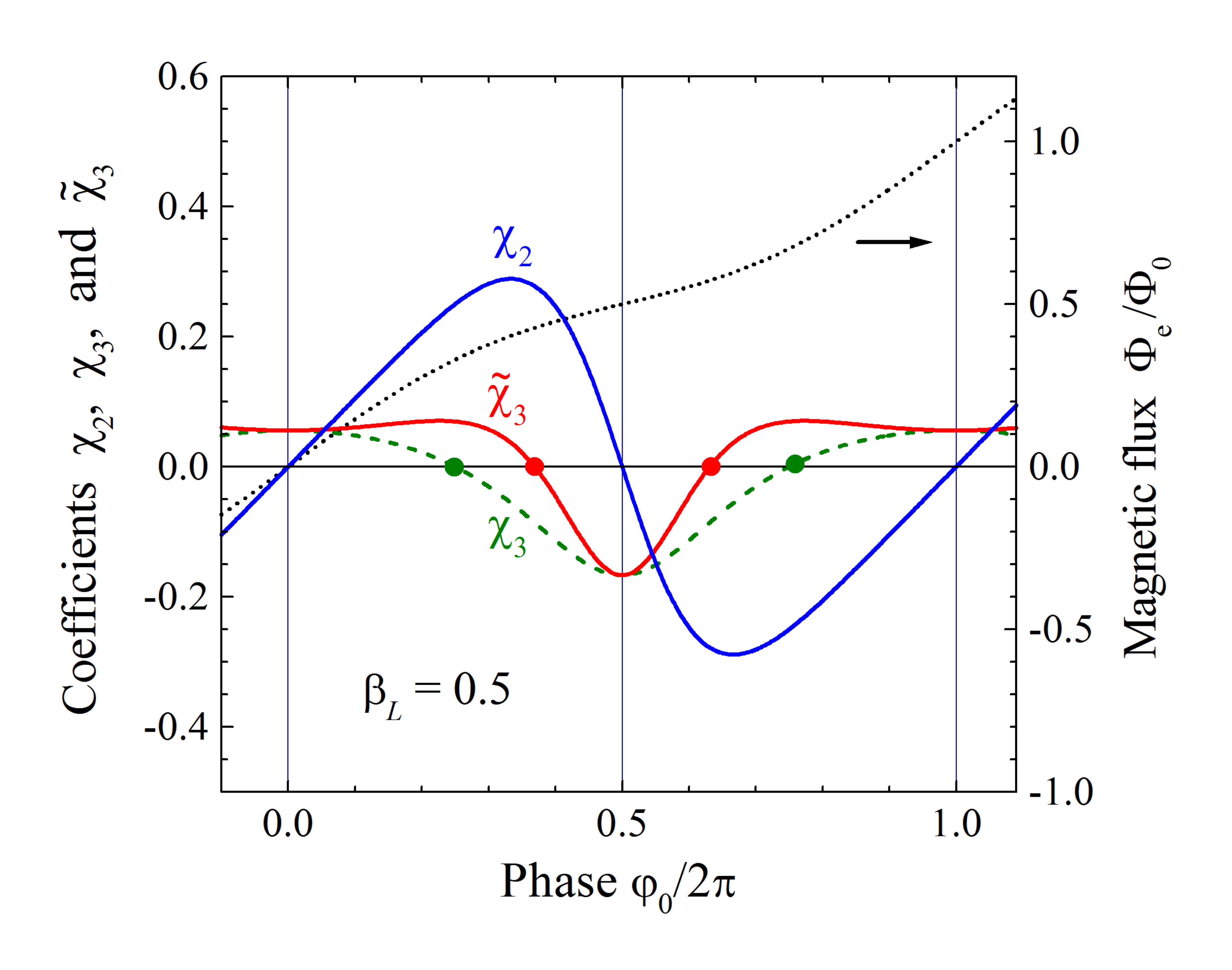}
\caption{Effective nonlinear coefficients $\chi_2$ (blue), $\chi_3$ (dashed green), and $\widetilde \chi_3$ (red) vs. phase bias $\varphi_0$ for $\beta_L=0.5$; since they are $2\pi$-periodic, only one period is shown. Due to corrections beyond the monochromatic approximation, the effective Kerr coefficient $\widetilde\chi_3$ is negative in a shorter range of $\varphi_0$-values around $\pi$ than $\chi_3$.
The phase bias $\varphi_0$ is determined by the external flux $\Phi_e$ (dotted line, right scale); for $\beta_L<1$ the dependence $\varphi_0(\Phi_e)$ is single-valued, and the whole range of phase bias $\varphi_0$ is accessible.
\label{fig:betagammatilde}}
\end{center}
\end{figure}

For the rf-SQUID-based circuit (PDBR-1b), the behavior of the effective Kerr nonlinearity $\widetilde\chi_3$ in \eqref{eq:gammatil} is illustrated in Fig.~\ref{fig:betagammatilde}. Similar to $\chi_2$, its sign varies with the phase bias $\varphi_0$, which can be controlled via the flux $\Phi_e$ applied to the SQUID loop. At the Kerr-free points, $\widetilde\chi_3=0$ (red dot in Fig.~\ref{fig:betagammatilde}), various phenomena based on pure three-wave mixing may be realized~\cite{Sivak2019,Miano2022,Khabipov_2022}.

\section{Other bifurcation detectors}

We now comment on the EOM for other bifurcation readout devices, such as the JBA (either in the conventional configuration with a dc-current-biased Josephson junction in Fig.~\ref{fig:EqvSchm}a, or in the rf-SQUID configuration in Fig.~\ref{fig:EqvSchm}b) and the PDBR-2 (based on a parametrically driven nonlinear resonator including a dc-SQUID shown in Fig.~\ref{fig:EqvSchm}c).
For the JBA, similar analysis yields an equation, reminiscent of Eq.~(\ref{eq:A1}),
\begin{equation}
\label{eq:A1-JBA}
\dot A_1 =
-\left(\theta-i\frac{\xi}{2}\right) A_1
+\frac{3i}{2}\left(\chi_3+\frac{10}{9}\chi_2^2\right)|A_1|^2 A_1
+ \frac{i}{2} \PPJ\,,
\end{equation}
where we keep only lowest-order terms, essential for the description of the resonance peak.
Here the drive term $i\PPJ/2$ is induced by near-resonant pumping, used in the operation of the JBA.
One can see that, as expected, the modified coefficient $\widetilde\chi_3$ is again given by Eq.~(\ref{eq:gammatil}) derived for the PDBR-1, since it does not contain any details of the drive.
Without this modification Eq.~(\ref{eq:A1-JBA}) coincides with (the low-order terms of) the
respective equation for the JBA in the monochromatic approximation~\cite{IthierThesis},
while the modification generally gives a substantial correction.

Both circuits in Figs.~\ref{fig:EqvSchm}a and \ref{fig:EqvSchm}b can also be operated in the JBA regime. In the latter case, the correction manifests itself in a shift of the Kerr-free points toward $\pm \pi$
and the corresponding narrowing of the interval with negative $\widetilde\chi_3$, see~Fig.~\ref{fig:betagammatilde}.
JBA with a negative Kerr nonlinearity was demonstrated~\cite{Khabipov2018} in a superconducting Nb coplanar-waveguide $\lambda/2$-resonator
with an rf-SQUID embedded in the central conductor~\cite{Khabipov_2022}.
In this case, the resonance curve exhibits a characteristic right slope.

Finally, we also consider the evolution equation for the PDBR-2 with a parametric driving $\propto \PPII x$,
\begin{equation}
\label{normalized_Eq:par}
\ddot{x} + x = \xi x -2\theta \dot{x} + \chi_2 x^2 + \chi_3 x^3 -\chi_4 x^4 +\PPII x\,\cos2\tau \,.
\end{equation}
In this case, analysis, similar to that performed above for PDBR-1, yields an EOM for the amplitude $A_1$ in the form
\begin{equation}
\label{eq:A1-par}
\dot A_1 =
-\left(\theta-i\frac{\xi}{2}\right) A_1
+\frac{i}{4}\PPII A_1^*
+\frac{3i}{2}\left(\chi_3+\frac{10}{9}\chi_2^2\right)|A_1|^2 A_1
-\frac{i}{32}\left(\chi_3+\frac{14}{3}\chi_2^2\right) \left[\PPII^*A_1^3+3\PPII|A_1|^2A_1^*\right].
\end{equation}
This equation was derived to the lowest orders in the amplitudes of the oscillations and the drive, $A_1$ and $\PPII$, respectively.
The second term ($\propto \PPII$) on the rhs of this equation induces parametric excitation similarly to the second term ($\propto \chi_2 \PPI$) on the rhs of Eq.~\eqref{eq:A1} originating from down-conversion of the drive in PDBR-1.
Note that Eq.~(\ref{eq:A1-par}) also satisfies the constraints, set by the Hamiltonian nature of the non-dissipative terms.

\section{Discussion}
In summary, we demonstrated the important role played by weak higher harmonics (with $n=0$ and
$|n|\ge2$) in the dynamics of the basic oscillations ($n=\pm1$) of a nonlinear resonator.
Nonlinearities create these higher harmonics from the basic oscillations and couple them back to the evolution
of the basic harmonic.
We demonstrated that the equations of motion for the basic harmonic are of Hamiltonian nature
(except the dissipative and noise terms), similar to the effective-potential description of the Kapitza inverted pendulum~\cite{LLMech}. Time-translation invariance strongly limits the functional form of the Hamiltonian.
The low-order terms in the Hamiltonian and dynamical equations
are given by Eq.~(\ref{eq:A1tilde}) for PDBR-1 (the period-doubling bifurcation readout with a current-biased junction, Fig.~\ref{fig:EqvSchm}a, or an rf-SQUID, Fig.~\ref{fig:EqvSchm}b), by Eq.~(\ref{eq:A1-par}) for PDBR-2 (PDBR in an dc-SQUID configuration of Fig.~\ref{fig:EqvSchm}c), and Eq.~(\ref{eq:A1-JBA}) for JBA (Josephson bifurcation amplifier), respectively.
These results allow for accurate analysis of the stationary states and transitions between them,
when control parameters of the circuit are varied. The relevant analysis of the Fokker-Planck equation and the switching process for PDBR-1, which defines the characteristics of this quantum detector, can be found in Ref.~\cite{pdbr}. This analysis demonstrated that the PDBR has properties comparable to those of the JBA, and for some parameter regimes exceeding those of the JBA.
We emphasize that the corrections beyond the monochromatic approximation that we found in this article, are essential for the quantitative description of various bifurcation quantum detectors, proposed and applied for readout of Josephson qubits.

We point out that nonlinear coefficients in these bifurcation readout circuits crucially depend on the bias point and can be efficiently tuned via the dc control current or external magnetic flux.
This flexibility may be used for improving sensitivity and other characteristics of the bifurcation readout devices. For example, remarkable properties of the rf-SQUID-based bifurcation circuit (PDBR-1b) allow one to tune the Kerr coefficient to a small value and ensure a sufficiently steep slope of the parametric resonance curve, which may notably improve resolution in reading out a qubit state.

We analyzed various bifurcation regimes using a lumped-element model of nonlinear resonant circuits. 
Experimentally, practical advantages, such as convenient control of the quality factor and a simple coplanar-waveguide design, were demonstrated in a cavity bifurcation amplifier (CBA)~\cite{Metcalfe2007}. CBA is based on a cavity-type superconducting microwave resonator with an embedded nonlinear element, which in its turn is integrated with a qubit. Its behavior is normally described within the monochromatic approximation by an equivalent lumped-element
circuit with effective parameters~\cite{Vijay}. This model, however, is valid only in the vicinity of the fundamental 
resonant frequency of the microwave-driven cavity. For higher oscillating modes, parameters of equivalent resonant circuits differ for each mode~\cite{Pozar-book}, 
and thus analysis of the effect of these modes on the CBA requires modifications of the developed approach. In particular, dynamical equations for the modes are coupled via nonlinear terms.
For cavity-based PDBR circuits, such as a coplanar-waveguide $\lambda/2$-resonator with an embedded rf-SQUID~\cite{Khabipov_2022}, analysis beyond the monochromatic (single-mode) approximation is particularly important because in this case the microwave drive is normally almost in resonance with the $\lambda$-mode.
Moreover, modifications of the resonator design toward strong coupling between the $\lambda/2$- and $\lambda$-modes~\cite{Khabipov_2022} enabled observation of the period-doubling regime at sufficiently weak microwave drive amplitudes.
We leave the analysis of this case for a subsequent investigation.

\begin{acknowledgments}
We are grateful to B.~Shteynas for his valuable contributions at an early stage of this work. We acknowledge valuable discussions with A.~Shnirman and D.~Zverevich.
\end{acknowledgments}

\bibliography{pdbr}

\end{document}